\documentclass{appolb}
\RequirePackage[T1]{fontenc}
\usepackage{graphicx}
\usepackage{hyperref}
\usepackage{amsmath}
\usepackage{caption}
\usepackage{subcaption}

\begin{document}
\title{Precise measurement of light-quark electroweak couplings at future colliders
\thanks{Presented at the XXX Cracow EPIPHANY Conference on Precision Physics at High Energy Colliders}%
}
\author{\underline{Krzysztof M\k{e}ka{\l}a}
\address{\hspace*{-3mm} Deutsches Elektronen-Synchrotron DESY, Notkestr. 85, 22607 Hamburg, Germany\\
Faculty of Physics, University of Warsaw, Pasteura 5, 02-093 Warsaw, Poland}
\\[3mm]
Daniel Jeans
\address{KEK, 1-1 Oho Tsukuba, Ibaraki 305-0801, Japan}
\\[3mm]
Junping Tian
\address{International Center for Elementary Particle Physics ICEPP, University of Tokyo, 7-3-1 Hongo, Bunkyo-ku, Tokyo 113-0033, Japan}
\\[3mm]
J\"urgen Reuter
\address{\hspace*{-3mm} Deutsches Elektronen-Synchrotron DESY, Notkestr. 85, 22607 Hamburg, Germany}
\\[3mm]
Aleksander Filip Żarnecki
\address{Faculty of Physics, University of Warsaw, Pasteura 5, 02-093 Warsaw, Poland}
}

\date{April 7, 2024}
\maketitle
\begin{abstract}
Electroweak Precision Measurements are stringent tests of the Standard Model and sensitive probes to New Physics. Accurate studies of the $Z$-boson couplings to the first-generation quarks could reveal potential discrepancies between the fundamental theory and experimental data. Future lepton colliders offering high statistics of $Z$ bosons would be an excellent tool to perform such a measurement based on comparison of radiative and non-radiative hadronic decays of the $Z$ boson. Due to the difference in quark charge, the relative contribution of the events with final-state radiation (FSR) directly reflects the ratio of up- and down-type quark decays. Such an analysis requires a proper distinction between photons coming from different sources, including initial-state radiation (ISR), FSR, parton showers and hadronisation. In our talk, we will show how to extract the values of the $Z$ couplings to quarks and present preliminary results of the analysis for ILC.
\end{abstract}
  
\section{Motivation}
The Standard Model (SM) of particle physics is the most successful theory describing interactions at the fundamental level. Despite its ability to explain experimental data from various experiments, several cosmological observations suggest that it is not the ultimate theory. Many convincing arguments, such as the density of dark matter, the baryon-antibaryon asymmetry or the very early history of the Universe, point towards the existence of a more fundamental model. Furthermore, the SM is a theory consistent with data only when a considerable number of internal parameters, including particle masses and couplings, is set to some specific values. Thus, precise measurements of the parameters are needed to probe the structure of the SM and possibly find any discrepancies revealing connections to more general models incorporating explanations for effects unpredicted by the SM.

Electroweak precision observables, including couplings of the SM fermions to photons, $Z$ and $W$ bosons, are among the best-measured parameters of this fundamental theory. The predictivity of the electroweak interactions at collider scales allows for comparing theoretical calculations with experimental data even at per-mille precision; for instance, the partial width of the $Z$-boson decaying into $b\bar{b}$ pairs is known to 0.3\%, while the partial width for the $c$-quark channel is constrained up to 1.7\%~\cite{ParticleDataGroup:2022pth}. On the other hand, the partial widths for the lighter quarks remain only poorly constrained due to the lack of proper tagging algorithms. Their values can however be extracted in another way, namely by studying radiative and non-radiative decays separately. Future Higgs factories operating at the $Z$-pole would offer production of numerous, $10^{9}$ -- $10^{12}$ $Z$ bosons making the measurement a sensitive probe of the quark-flavour universality~\cite{deBlas:2022ofj}. In this talk, we presented prospects for measuring the light-quark couplings at colliders of this type.

\section{Measurement idea}
\label{sec:idea}
The measurement has already been performed at LEP~\cite{DELPHI:1991utk, L3:1992kcg, L3:1992ukp, DELPHI:1995okj, OPAL:2003gdu}. The general idea is based on the fact that up- and down-type quarks differ in electric charge. The coupling strength of the $Z$ boson to a given fermion $f$ is conventionally defined as
\begin{equation}
    c_f = v^2_f + a^2_f,
\end{equation}
where $v_f$ ($a_f$) is the vector (axial) coupling. They are determined in the SM by the third component of the fermion weak isospin and its charge:
\begin{equation}
    v_f = 2I_{3,f}-4Q_f \sin^2{\theta_W}, \quad \quad a_f = 2I_{3,f},
\end{equation}
where $\theta_W$ is the weak mixing angle.
The total width of the $Z$ boson to hadrons, $\Gamma_{had}$, reads:
\begin{equation}
    \Gamma_{had} = N_c \frac{G_\mu M^3_Z}{24\pi \sqrt{2}}\left(1 + \frac{\alpha_s}{\pi}+ \mathcal{O}(\alpha_s^2)\right) (3c_{d} + 2c_{u}),
\end{equation}
where $N_c$ is the number of colours, $G_\mu$ is the Fermi constant, $M_Z$ is the mass of the $Z$ boson, $\alpha_s$ is the strong coupling constant and $c_d$ ($c_u$) is the coupling to down-type (up-type) quarks~\cite{L3:1992kcg,OPAL:2003gdu}. The total width to radiative hadronic decays, $\Gamma_{had+\gamma}$, is given by
\begin{equation}
    \Gamma_{had+\gamma} = N_c \frac{G_\mu M^3_Z}{24\pi \sqrt{2}} f(y_{cut})\frac{\alpha}{2\pi} \left(3q_d^2 c_{d} + 2q_u^2c_{u}\right),
\end{equation}
where $f(y_{cut})$ is a form factor depending on the arbitrary parameter $y_{cut}$ incorporating the isolation criteria for photons, $\alpha$ is the electromagnetic coupling constant and $q_d$ ($q_u$) is the electric charge of down-type (up-type) quarks. Since $q_d \ne q_u$, the expressions for the radiative and the total hadronic widths include different coupling combinations and the couplings of the up- and down-type quarks can be disentangled.

Specifically, given excellent heavy-flavour tagging envisioned for future $e^+e^-$ colliders~\cite{Liang:2023yyi}, one may assume for simplicity that $b\bar{b}$ and $c\bar{c}$ events can be fully separated from the light-flavour production data. Then, the ``light'' hadronic cross section for $Z \to q\bar{q}$, $q = u, d, s$, can be expressed as
\begin{equation}
    \sigma_{_{Z\to lhad}} = \mathcal{C}_1 \cdot (2 c_d + c_u), 
\end{equation}
where $\mathcal{C}_1$ is a constant. Similarly, the radiative hadronic cross section is given by 
\begin{equation}
       \sigma_{_{Z\to lhad+\gamma}} = \mathcal{C}_2 \cdot (c_d + 2 c_u),
\end{equation}
where $\mathcal{C}_2$ is another constant (if the reconstruction parameters are fixed to some particular values). By fitting the experimental data to theoretical calculations and
simulations,
the values of $\mathcal{C}_1$ and $\mathcal{C}_2$ can be extracted, and one can disentangle $c_d$ and $c_u$.

\begin{figure}
    \centering
    \includegraphics[width=0.7\textwidth]{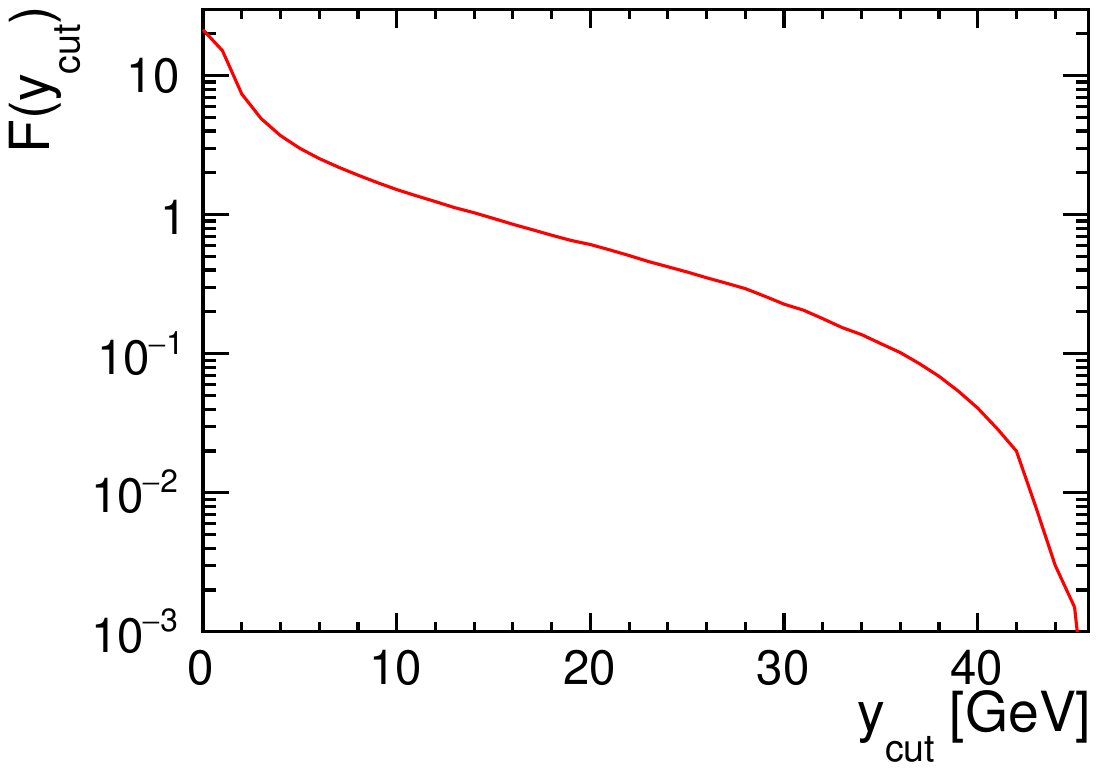}
    \caption{The form factor $\textnormal{F}(y_{cut})$ as a function of the photon transverse momentum with respect to the closest quark axis. The results were obtained in \textsc{Whizard 3.1}.}
    \label{fig:Fy}
\end{figure}

Naturally, the two constants are connected and the radiative cross section can be rewritten as: 
\begin{equation}
    \sigma_{_{Z\to lhad+\gamma}} = \sigma_{_{Z\to lhad}} \cdot \frac{c_d + 2 c_u}{2c_d + c_u} \cdot \frac{\alpha} {\pi}\cdot \textnormal{F}(y_{cut}).
\end{equation}
The form factor $\textnormal{F}(y_{cut})$ may depend on an arbitrarily chosen cut parameter describing the separation between final-state quarks and emitted photons. The function can be extracted from precise theoretical calculations at fixed order. For illustrative purposes, the leading-order value has been obtained in \textsc{Whizard 3.1}~\cite{Kilian_2011, moretti2001omega} and is shown in Figure \ref{fig:Fy} where $y_{cut}$ is taken to be the photon transverse momentum with respect to the closest quark axis.

\section{Simulation of events}
The measurement relies on precise Monte Carlo generation of data samples. The process of $e^+e^- \to q\bar{q}$ is often perceived as a benchmark point for testing Monte Carlo tools and typically does not pose a challenge for the generators. However, the requirement of reconstruction of isolated photons introduces several complications. One has to consider not only the final-state photons (FSR) from Matrix Elements (ME) but also to model properly initial-state radiation (ISR), match fixed-order calculations to parton showers and separate photons occurring from hadronisation. As a recipe, one can generate data samples using fixed-order ME calculations, with exclusive emissions of hard photons, and match them with ISR structure function and FSR showers accounting for collinear and soft emissions. A similar problem has been tackled in \cite{Kalinowski:2020lhp} where dark-matter production at lepton colliders has been considered. A matching procedure for simulating photons using both ME calculations (for hard emissions) and ISR structure function (for soft emissions) has been developed for \textsc{Whizard}~\cite{Kilian_2011,moretti2001omega}, a Monte Carlo generator incorporating many features important for future lepton colliders, including beam polarisation, beamstrahlung and ISR spectra. We followed and extended this approach to account for effects present for hadronic final states. For this purpose, the internal interface of \textsc{Whizard} to \textsc{Pythia\,6}~\cite{Sjostrand:2006za} was employed to include parton showers and hadronisation. In future, we plan to compare our results with those from \textsc{Pythia 8}~\cite{Sjostrand:2007gs}, \textsc{Herwig 7}~\cite{Bellm:2015jjp} and \textsc{KKMC 5}~\cite{Arbuzov:2020coe} to assess systematic uncertainties coming from the Monte Carlo modelling.

\begin{figure}
     \centering
     \begin{subfigure}[b]{0.45\textwidth}
         \centering
         \includegraphics[width=\textwidth]{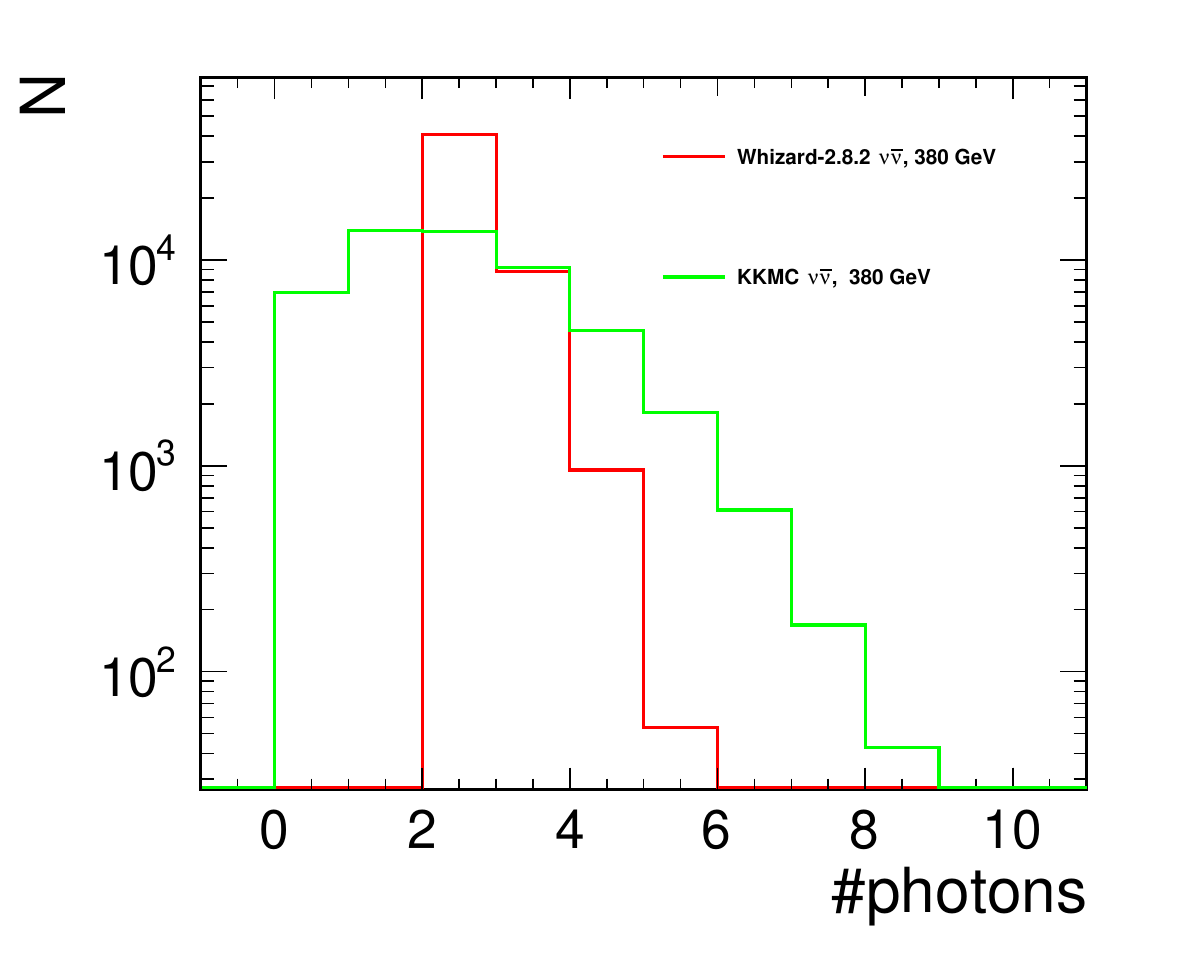}
         \caption{All generated photons}
     \end{subfigure}
     \hfill
     \begin{subfigure}[b]{0.45\textwidth}
         \centering
         \includegraphics[width=\textwidth]{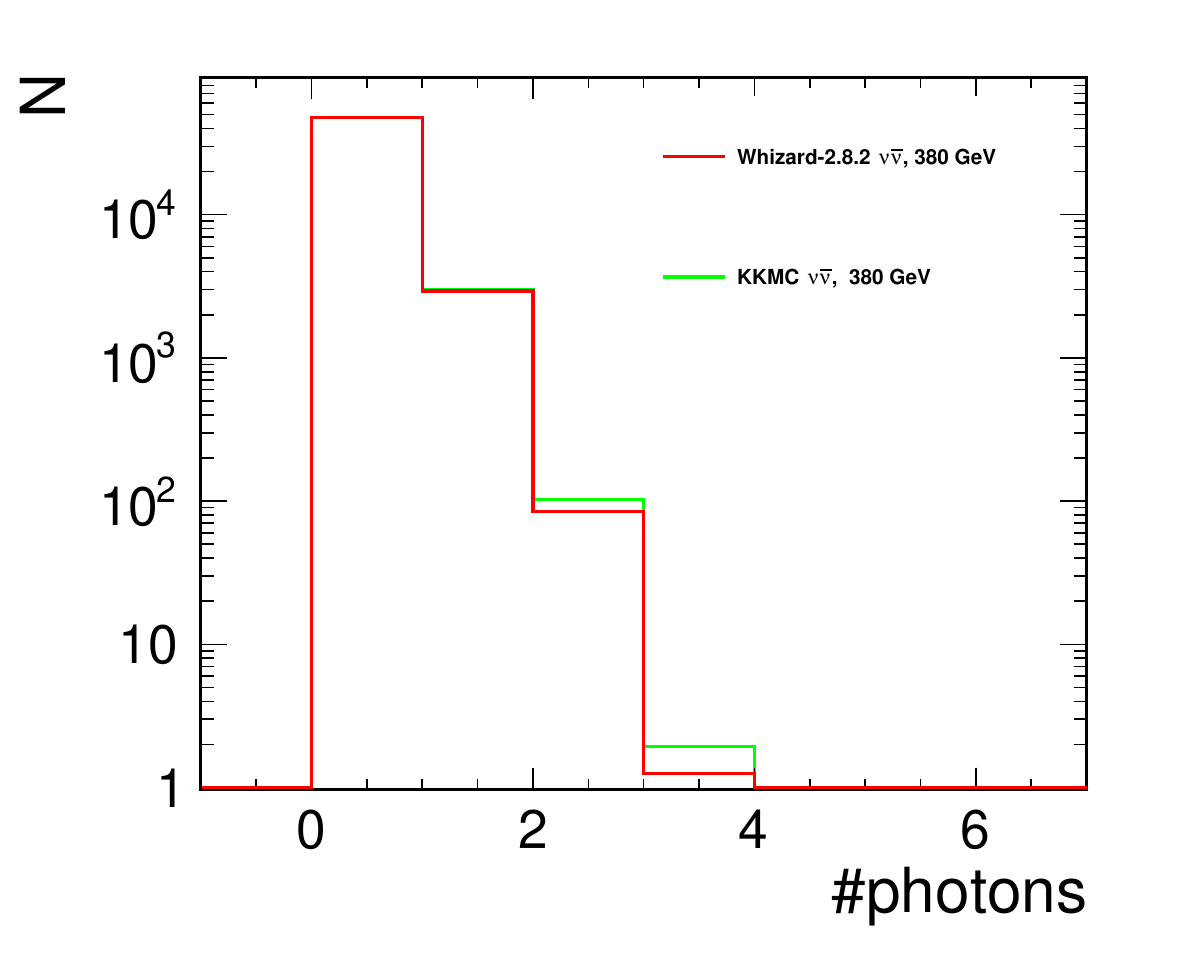}
         \caption{Hard photons only}
     \end{subfigure}
        \caption{Distributions of the photon multiplicities for the process of $e^+e^- \to \nu \bar{\nu}$ at collision energy of 380 GeV simulated in \textsc{Whizard} and \textsc{KKMC} (the Z-pole run was not considered in that study~\cite{Kalinowski:2020lhp}). The left plot shows the multiplicity for all photons generated by the MC codes and the right one the multiplicity of the photons passing selection criteria (``hard photons'', observable in the detector). As the final state is electrically neutral, only initial-state photons are expected in this process. The distributions are normalised to the number of events expected for an integrated luminosity of 1 fb$^{-1}$.}
        \label{fig:KKMC}
\end{figure}

The matching in \cite{Kalinowski:2020lhp} relied on two variables, $q_-$ and $q_+$, defined as:
\begin{align*}
    q_- = \sqrt{4E_0 E_\gamma}\sin \frac{\theta_\gamma}{2},\\
    q_+ = \sqrt{4E_0 E_\gamma}\cos \frac{\theta_\gamma}{2},
\end{align*}
where $E_0$ is the nominal beam energy, $E_\gamma$ is the energy of the emitted photon and $\theta_\gamma$ is its emission angle. The variable $q_-$ ($q_+$) corresponds to the virtuality of a single photon emitted from the electron (positron) line. The procedure assumes that all the photons with $q_\pm > 1$ GeV and $E_\gamma > 1$ GeV are generated via fixed-order calculations while all the softer emissions are taken into account via the built-in ISR structure function in \textsc{Whizard}. As shown in Figure \ref{fig:KKMC} taken from \cite{Kalinowski:2020lhp}, the procedure allows for restoring similar photon multiplicities as those obtained in \textsc{KKMC}~\cite{Jadach:1999vf}, a Monte Carlo code for $e^+e^- \to f\bar{f}$ employing the YFS exponentiation~\cite{Yennie:1961ad} for exclusive photon emissions. 

\begin{figure}
    \centering
    \includegraphics[width=0.7\textwidth]{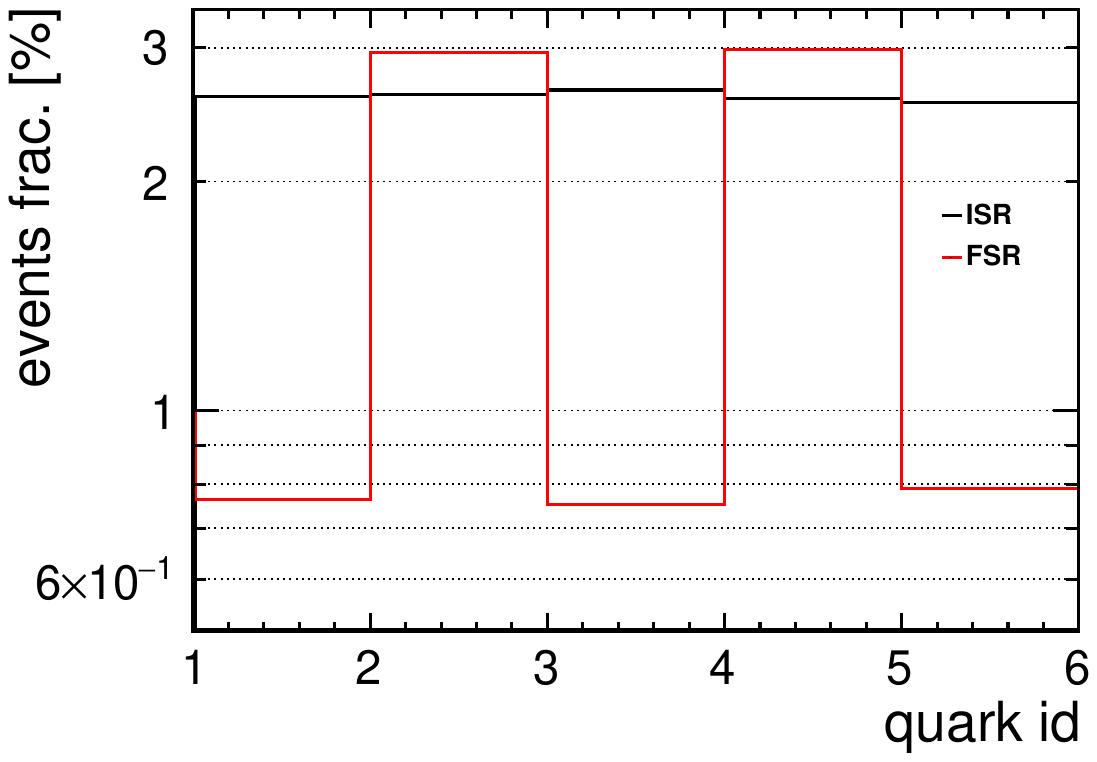}
    \caption{Fraction of events rejected by the matching procedure for different quark flavours (normalised to the cross section per flavour) for ISR (black) and FSR (red)}
    \label{fig:rejection}
\end{figure}

The procedure described above has been developed for dark matter pair-production, \textit{i.e.} for the case when final-state particles have been assumed to be chargeless (electrically neutral and colourless). When considering hadronic Z decays, QCD and QED showers need to be included for the hadronic final state and thus, the procedure had to be extended. For the separation of \textit{hard} and \textit{soft} final state radiation domains, we use an additional criterion based on the invariant mass of the photon-jet pairs, $M_{\gamma\, j}$. The \textit{hard} photons, \textit{i.e.} those fulfilling all the criteria simultaneously:
\begin{itemize}\setlength\itemsep{-0.2em}
    \item $q_\pm > 1$ GeV,
    \item $E_\gamma > 1$ GeV,
    \item $M_{\gamma\, j_{1,2}} > 1$ GeV,
\end{itemize}
were generated using fixed-order ME calculations and \textit{soft} photons (not meeting at least one of the criteria) were simulated via the internal structure function (for ISR) and the \textsc{Pythia6} parton shower (for FSR). Events with at least one ISR or FSR photon passing \textit{hard} photon selection were rejected (the matching procedure was established to avoid double counting). In Figure~\ref{fig:rejection}, we show the fraction of rejected events for different quark flavours (normalised to the cross section per flavour) for ISR and FSR. As expected, the rejection efficiency for ISR does not depend on the quark flavour while the rejection efficiency for FSR distinguishes up- and down-type quarks and differs by a factor of 4 between the two cases, which corresponds to the difference in the quark charge squared.

\section{Analysis procedure}
As explained in the previous section, measurable photons originate from different processes, including ISR, FSR and hadronisation. The matching procedure developed for this study allows for removing double counting of photons generated explicitly at fixed order and those coming from structure functions and showers. However, the separation of the FSR photons, providing insight into quark charges and thus, opening the path towards flavour tagging, has to be handled at the analysis level to emulate physical conditions of future experiments. At the $Z$-pole, the contribution from ISR is strongly suppressed due to the small phase space for photon emissions. However, the impact of photons coming from hadronic decays is non-negligible and forms the main part of the background in this analysis. 

To illustrate the issue, we generated 0- and 1-ME-photon samples in \textsc{Whizard 3.1} according to the matching procedure described above. Higher photon multiplicities have been neglected for now, as the cross section for the 2-ME-photon sample is about 30 times smaller than that for the 1-ME-photon sample. We performed fast detector simulation with \textsc{Delphes 3.5}~\cite{deFavereau:2013fsa} using the default \textit{ILCgen} cards employing the Durham algorithm~\cite{Dokshitzer:1997in} in the exclusive two-jet mode. For analysis, we accepted only two-jet events with exactly one reconstructed photon with transverse momentum $p^T_\gamma > 2$ GeV and pseudorapidity $|\eta| < 2.5$. More stringent cuts can help reject photons coming from hadronisation and their optimisation has been left for future studies.

\begin{figure}
    \centering
    \includegraphics[width=0.7\textwidth]{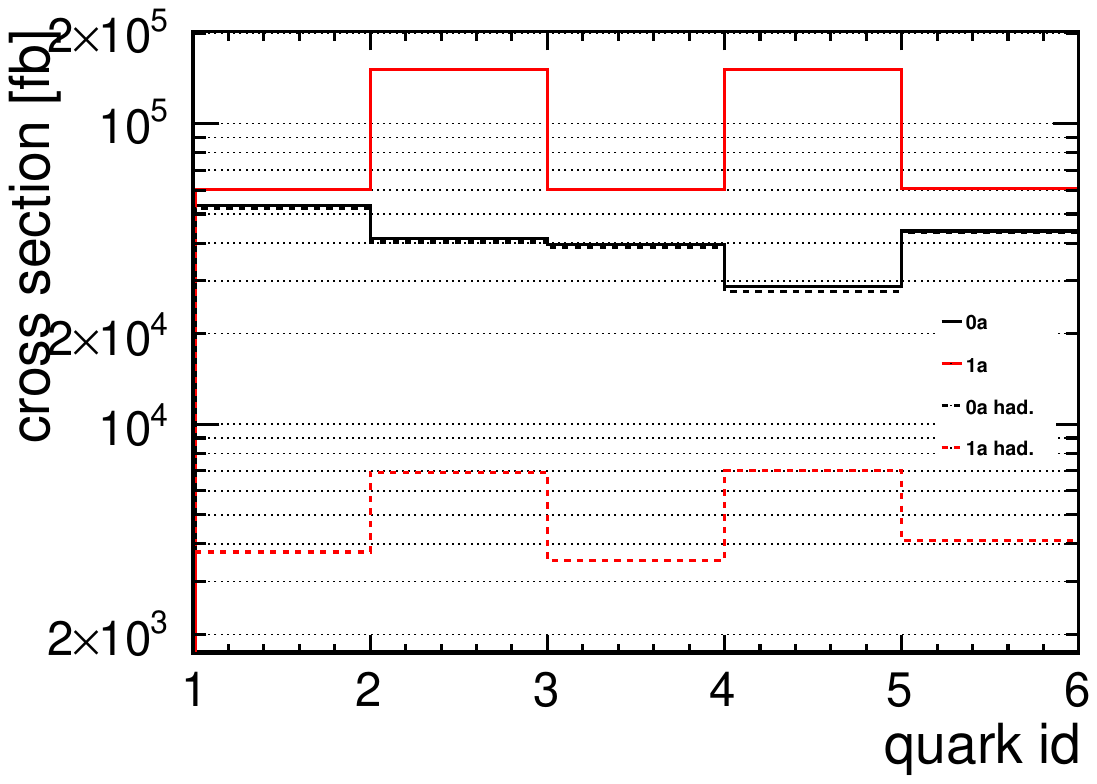}
    \caption{Cross sections for 0- (black) and 1-ME-photon (red) samples for different quark flavours. The solid lines indicate all the reconstructed events while the dashed lines stand for the photons coming from hadron decays only}
    \label{fig:hadro}
\end{figure}

In Figure \ref{fig:hadro}, we compare cross sections for 0- and 1-ME-photon samples taking into account photons coming from hadron decays, as well as all the reconstructed photons. The comparison clearly shows that photons in the 0-ME-photon sample originate entirely from decays of neutral hadrons, while in the 1-ME-photon sample, they form only about 10\% of all the photons. The composition of the hadron-decay background for different quark flavours for the 0-ME-photon sample is presented in Table \ref{tab:hadr}. For all the flavours, $\pi^0$ decays are dominant; another important contribution comes from $\eta$ decays. For the heavy quarks, one also has to include $D$ and $B$ meson decays.

\begin{table}
\centering
\begin{tabular}{c|c|c|c|c|c} 

{[}\%]   & $d$   & $u$   & $s$   & $c$   & $b$   \\ \hline
$\pi^0$       & 94  & 94  & 94  & 93  & 88  \\ \hline
$\eta$        & 4.5 & 4.5 & 4.3 & 3.7 & 3.6 \\ \hline
$D$ mesons & -   & -   & -   & 1.9 & 2.0   \\ \hline
$B$ mesons & -   & -   & -   & -   & 5.6 \\ 
\end{tabular}
\caption{Composition of the hadron-decay background for different quark flavours for the 0-ME-photon sample, based on the hadronisation model implemented in \textsc{Pythia 6}.}
\label{tab:hadr}
\end{table}

The severe background coming from hadron decays points towards the necessity of elaborated studies not only on the analysis cuts but also on the dedicated photon reconstruction criteria. As elucidated in Section \ref{sec:idea}, the isolation parameter $y_{cut}$ can be chosen arbitrarily, for instance, to minimise the systematic error of the measurement. We plan to optimise the reconstruction criteria in future.

\section{Conclusions and outlook}
Future $e^+e^-$ colliders operating at the $Z$-pole will effectively constrain the Standard Model parameters. Thanks to very high data statistics, they can allow for performing precision measurements of the $Z$-boson couplings to fermions, including those of light quarks. In our study, we established a dedicated generation procedure accounting for photons coming from different sources, including ISR, FSR, hadronisation and hadron decays. We performed preliminary studies on the experimental cuts and their efficiency. In the next steps, we plan to study photon isolation criteria, which are crucial for reducing the background originating from decays of neutral particles, and the uncertainty coming from Monte Carlo simulation. The ultimate goal of the study is to estimate the uncertainty of the measurement at different future machines depending on the reconstruction criteria and experimental cuts.

\section*{Acknowledgements}

The work was supported by the National Science Centre (Poland) under OPUS research project no.~2021/43/B/ST2/01778 and the Deutsche
Forschungsgemeinschaft (Germany) under
Germany's Excellence Strategy-EXC 2121 ``Quantum Universe''-390833306. Furthermore, we acknowledge support from
the COMETA COST Action CA22130 and from the International Center for Elementary Particle Physics (ICEPP), the University of Tokyo. 

\bibliographystyle{unsrt}
\bibliography{bibliography}

\end{document}